\documentclass[aps,graphicx,12pt]{revtex4}
\usepackage{amssymb}
\usepackage{amsmath}
\usepackage{amscd}
\usepackage{subfigure}
\usepackage{graphicx}

\begin{document}

\title{Generation of $N$-atom $W$-class states in spatially separated
cavities}
\author{Mei Lu$^{1}$}
\author{Yan Xia$^{1,}$\footnote{E-mail: xia-208@163.com}}
\author{Jie Song$^{2}$}
\author{Nguyen Ba An$^{3,}$\footnote{E-mail: nban@iop.vast.ac.vn}}

\affiliation{$^{1}$Department of Physics, Fuzhou University, Fuzhou
350002, China\\$^{2}$Department of Physics, Harbin Institute of
Technology, Harbin, Heilongjiang 150001, China\\$^{3}$Center for
Theoretical Physics, Institute of Physics, 10 Dao Tan, Hanoi,
Vietnam}

\begin{abstract}
We propose a feasible and efficient scheme to generate $N$-atom
$W$-class states in spatially separated cavities without using any
classical driving pulses. We adopt the model in which the couplings
between different atoms are mediated only by virtual excitations
of the cavity and fiber fields, so the scheme is insensitive to the cavity decay
and fiber photon leakage. We carry out both theoretical
investigation in a decoherence-free subspace and numerical
calculation accounting for decoherence due to the atomic spontaneous
emission as well as the decay of cavity and fiber modes. The
theoretical and numerical results agree in the large atom-cavity
detuning regime. Our scheme proves to be useful in scalable
distributed quantum networks.
\end{abstract}

\maketitle

\section{INTRODUCTION}

Entanglement, a fundamental feature in quantum mechanics, is a key resource
for quantum information processing (QIP) \cite{n-hjk53,prl-sbz92}, such as
quantum teleportation \cite{prl-chb70}, quantum dense coding \cite{prl-chb69}%
, quantum cryptography \cite{prl-ake67} and quantum computation \cite{n-dg02}%
. If a composite system is entangled, the whole system cannot be split into
independent subsystems. Typical entangled states are the Bell states \cite
{p-jsb95}, the Greenberger-Horne-Zeilinger-class states \cite
{ajp-dmg31,apl-xy08} and the $W$-class states \cite{pra-wd14}. A state is
called an $N$-qubit $W$-class state if it is of the form $x_{1}|10\cdots
0\rangle +x_{2}|01\cdots 0\rangle +\cdots +x_{N}|00\cdots 1\rangle $ with $%
\sum_{n=1}^{N}|x_{n}|^{2}=1$ and ${|0\rangle ,|1\rangle }$ being two
orthonormal vectors in the two-dimensional Hilbert space of the qubit. The $%
N $-qubit $W$ state corresponds to $x_{1}=x_{2}=\cdots =x_{N}=1/\sqrt{N}.$
Compared with other types of entangled states, $W$-class state constitutes a
very important family of states possessing a high degree of robustness
against the qubit loss as they maintain some entanglement when more than two
qubits remained \cite{prl-hjb10,njp-jj36}. Furthermore, deterministic
protocols for teleportation and superdense coding \cite{pra-pa20} have been
designed by utilizing $W$-class entanglements \cite{qipxww31}. An
asymmetrical $N$-partite $W$-class state is also an essential quantum
channel for quantum information splitting \cite{prl-sbz03,ctp-hzw65} and can
be converted to another $N$-partite $W$-class state via local operations and
classical communications \cite{pra-wc14}. Therefore, the generation of $N$%
-partite $W$-class states has proved to be an urgent task for the QIP.

However, to our knowledge, there are few studies for the given operationally
experimental configurations with multipartite entanglement classes. Bastin
\emph{et al.} proposed an experimental setup to produce arbitrary symmetric
long-lived multiqubit $W$ states in the internal ground levels of photon
emitters \cite{prl-tb01}. An and Wang \textit{et al}. presented protocols to
generate $N$-party $W$-class states in a single optical microcavity \cite
{an,qipxww31}. The $W$-class states in the above protocols are generated
locally. For the distributed QIP, Pellizzari \cite{prl-tp42} first suggested
a scheme to realize the reliable transfer of quantum information between two
distant cavities connected by an optical fiber in 1997, providing an
effective tool for long-distant quantum communication schemes in recent
years \cite{prl-as03,epl-js01,pra-sbz27,pra-zqy24,pra-xyl30,pra-js08}. Here,
by using a single-mode integrated optical $1\times N$ star coupler \cite
{epjd-lts23,apl-tf49,jap-th36,a0-lby68}, we propose a scheme to generate $N$%
-atom $W$-class states in a distributed network, which is nonlocally
correlative \cite{pra-as06} even in the presence of noise. Another distinct
feature of our scheme is that all the bosonic field modes are only virtually
populated to overcome the decoherence caused by cavity decays and fiber
photon leakages. The excitation exchange among the atoms is caused by the
dispersive coupling between the atoms and multiple delocalized bosonic field
modes. Therefore, the atom-fiber-cavity system reduces to an effective model
that couples only the atomic states while suppresses the states containing
real bosonic modes. In addition, no classical pulses are needed so that the
scheme is easy to operate. All these features make the scheme very promising
for the generation of $N$-atom $W$-class states in spatially separated
cavities.

\section{THE\ MODEL}

We consider $N\ (N\geq 3)$ identical atoms trapped in $N$ distant cavities
which are connected by a single-mode integrated optical $1\times N$  star
coupler \cite{apl-tf49,jap-fh36,ao-lby68}, as shown in Fig. 1(a). The
optical star coupler is made up of $N$ identical optical fiber channels and
only one resonant field mode interacts simultaneously with $N$ cavity modes
with a (real) coupling constant $\nu .$ Each atom is a two-level one playing
the role of a qubit with $|0\rangle \equiv |g\rangle ,$ the ground state,
and $|1\rangle \equiv |e\rangle ,$ the excited state. The atomic transition
frequency $\Omega $ is detuned from the cavity mode frequency $\omega $ by a
certain amount $\Delta =\Omega -\omega ,$ as shown in Fig. 1(b). Thus, the
atomic transition $|g\rangle \leftrightarrow |e\rangle $ is dispersively
coupled to the corresponding cavity mode with a (real) coupling constant $f.$
The interaction Hamiltonian of the whole atom-cavity-fiber system under the
rotating-wave approximation can be written as $(\hbar =1)$
\begin{equation}
H=H_{1}+H_{2},  \label{1}
\end{equation}
\begin{equation}
H_{1}=\sum_{l=1}^{N}\nu (a_{l}^{\dagger }b+b^{\dagger }a_{l}),  \label{2}
\end{equation}
\begin{equation}
H_{2}=\sum_{l=1}^{N}f(a_{l}^{\dagger }S_{l}^{-}e^{-i\Delta
t}+S_{l}^{+}a_{l}e^{i\Delta t}),  \label{3}
\end{equation}
where $a_{l}^{\dagger }$ $(a_{l})$ is the creation (annihilation) operator
of the $l$th cavity mode, $b^{\dagger }$ $(b)$ is the creation
(annihilation) operator of the fiber mode and $S_{l}^{+}=|e_{l}\rangle
\langle g_{l}|$ $(S_{l}^{-}=|g_{l}\rangle \langle e_{l}|)$ denotes the
rasing (lowering) operator of the $l$th atom.

We introduce new bosonic operators $c_{\alpha }$ defined by a linear
superposition of $a_{l}\ (l=1,2,...,N)$ and $b$
\begin{equation}
c_{\alpha }=\sum_{l=1}^{N}t_{\alpha ,l}a_{l}+t_{\alpha ,N+1}b,  \label{4}
\end{equation}
where $t_{\alpha ,\beta }$ with $\alpha ,\beta \in \{1,2,...,N+1\}$ are the
elements of a $(N+1)\times (N+1)$ real unitary matrix $T$ of the form
\begin{equation}
T=\left(
\begin{tabular}{cccccccc}
$\frac{\sqrt{N-1}}{\sqrt{N}}$ & $\frac{-1}{\sqrt{N(N-1)}}$ & $\frac{-1}{%
\sqrt{N(N-1)}}$ & $\cdots $ & $\frac{-1}{\sqrt{N(N-1)}}$ & $\frac{-1}{\sqrt{%
N(N-1)}}$ & $\frac{-1}{\sqrt{N(N-1)}}$ & $0$ \\
$0$ & $\frac{\sqrt{N-2}}{\sqrt{N-1}}$ & $\frac{-1}{\sqrt{(N-1)(N-2)}}$ & $%
\cdots $ & $\frac{-1}{\sqrt{(N-1)(N-2)}}$ & $\frac{-1}{\sqrt{(N-1)(N-2)}}$ &
$\frac{-1}{\sqrt{(N-1)(N-2)}}$ & $0$ \\
$0$ & $0$ & $\frac{\sqrt{N-3}}{\sqrt{N-2}}$ & $\cdots $ & $\frac{-1}{\sqrt{%
(N-2)(N-3)}}$ & $\frac{-1}{\sqrt{(N-2)(N-3)}}$ & $\frac{-1}{\sqrt{(N-2)(N-3)}%
}$ & $0$ \\
$\vdots $ & $\vdots $ & $\vdots $ & $\ddots $ & $\vdots $ & $\vdots $ & $%
\vdots $ & $\vdots $ \\
$0$ & $0$ & $0$ & $\cdots $ & $\frac{\sqrt{2}}{\sqrt{3}}$ & $\frac{-1}{\sqrt{%
3\times 2}}$ & $\frac{-1}{\sqrt{3\times 2}}$ & $0$ \\
$0$ & $0$ & $0$ & $\cdots $ & $0$ & $\frac{1}{\sqrt{2}}$ & $\frac{-1}{\sqrt{%
2\times 1}}$ & $0$ \\
$\frac{-1}{\sqrt{2N}}$ & $\frac{-1}{\sqrt{2N}}$ & $\frac{-1}{\sqrt{2N}}$ & $%
\cdots $ & $\frac{-1}{\sqrt{2N}}$ & $\frac{-1}{\sqrt{2N}}$ & $\frac{-1}{%
\sqrt{2N}}$ & $\frac{1}{\sqrt{2}}$ \\
$\frac{1}{\sqrt{2N}}$ & $\frac{1}{\sqrt{2N}}$ & $\frac{1}{\sqrt{2N}}$ & $%
\cdots $ & $\frac{1}{\sqrt{2N}}$ & $\frac{1}{\sqrt{2N}}$ & $\frac{1}{\sqrt{2N%
}}$ & $\frac{1}{\sqrt{2}}$%
\end{tabular}
\ \ \right) .  \label{5}
\end{equation}
The inverse transformations of Eq. (\ref{4}) are
\begin{eqnarray}
a_{l} &=&\sum_{\alpha =1}^{N+1}\chi _{l,\alpha }c_{\alpha },  \label{6} \\
b &=&\sum_{\alpha =1}^{N+1}\chi _{N+1,\alpha }c_{\alpha },  \label{7}
\end{eqnarray}
where $\chi _{\alpha ,\beta }$ are the elements of a $(N+1)\times (N+1)$
real unitary matrix
\begin{equation}
X=T^{-1}=T^{T}.  \label{8}
\end{equation}
In terms of the new delocalized bosonic operators in Eq. (\ref{4}), $H_{1}$
and $H_{2}$ read
\begin{equation}
H_{1}=-\sqrt{N}\nu \left( c_{N}^{\dagger }c_{N}-c_{N+1}^{\dagger
}c_{N+1}\right)   \label{9}
\end{equation}
and
\begin{equation}
H_{2}=\sum_{l=1}^{N}\sum_{\alpha =1}^{N+1}f\chi _{l,\alpha }\left(
S_{l}^{+}c_{\alpha }e^{i\Delta t}+c_{\alpha }^{\dagger }S_{l}^{-}e^{-i\Delta
t}\right) .  \label{10}
\end{equation}
Switching to the interaction representation, $\mathcal{H}=\mathcal{H}_{0}+%
\mathcal{H}_{int},$ with $\mathcal{H}_{0}=H_{1},$ we have
\begin{equation}
\mathcal{H}_{int}=\sum_{l=1}^{N}\sum_{\alpha =1}^{N+1}f\chi _{l,\alpha
}\left( S_{l}^{+}c_{\alpha }e^{i\Delta _{\alpha }t}+c_{\alpha }^{\dagger
}S_{l}^{-}e^{-i\Delta _{\alpha }t}\right) ,  \label{11}
\end{equation}
where
\begin{equation}
\Delta _{\alpha }=\left\{
\begin{tabular}{ccc}
$\Delta $ & for & $\alpha =1,2,...,N-1$ \\
$\Delta +\sqrt{N}\nu $ & for & $\alpha =N$ \\
$\Delta -\sqrt{N}\nu $ & for & $\alpha =N+1$%
\end{tabular}
\right. .  \label{12}
\end{equation}

We assume that all the cavities and the fibers are empty and only one atom
is excited initially. Then, in the large detuning regime: $\Delta ,$ $%
|\Delta \pm \sqrt{N}\nu |\gg f,$ the atoms are forbidden to exchange energy
with the bosonic fields, but they can exchange energy with each other via
virtual field modes. So during the system's evolution no real bosonic modes
appear at all, but the atomic excitation can still propagate from atom to
atom. The underlying dynamics is thus governed by the effective interaction
Hamiltonian
\begin{equation}
\mathcal{H}_{eff}=\sum_{l,m=1}^{N}\xi _{lm}S_{l}^{+}S_{m}^{-},  \label{13}
\end{equation}
with
\begin{equation}
\xi _{lm}=f^{2}\sum_{\alpha =1}^{N+1}\frac{\chi _{l,\alpha }\chi _{m,\alpha }%
}{\Delta _{\alpha }}.  \label{14}
\end{equation}
Using the equalities $\chi _{l,\alpha }=t_{\alpha ,l}$ due to Eq. (\ref{8}),
the unitarity of $T:$ $\sum_{\alpha =1}^{N+1}t_{\alpha ,\beta }t_{\alpha
,\delta }=\delta _{\beta \delta },$ and the properties $-t_{N,\beta
}=t_{N+1,\beta }=1/\sqrt{2N}$ for $\beta \in \{1,2,...,N\},$ we can verify
that
\begin{equation}
\xi _{lm}=\left\{
\begin{tabular}{ccc}
$\frac{f^{2}}{N}\left( \frac{N-1}{\Delta }+\frac{\Delta }{\Delta ^{2}-N\nu
^{2}}\right) =\xi _{N}$ & for & $l=m$ \\
$-\frac{f^{2}}{N}\left( \frac{1}{\Delta }-\frac{\Delta }{\Delta ^{2}-N\nu
^{2}}\right) =-\eta _{N}$ & for & $l\neq m$%
\end{tabular}
\right. .  \label{15}
\end{equation}

Now we turn to the generation of $N$-atom $W$-class states via the effective
interaction Hamiltonian in Eq. (\ref{13}). In the subspace having only one
excited atom and no real bosonic modes, the atoms' state at any time $t$ can
be represented by a linear superposition of $N$ basic states $\{|\phi
_{n}\rangle ;n=1,2,...,N\}$ as
\begin{equation}
\left| \Phi _{N}(t)\right\rangle =\sum_{n=1}^{N}C_{n}^{(N)}(t)|\phi
_{n}\rangle ,  \label{16}
\end{equation}
where $|\phi _{n}\rangle =|...e_{n}...\rangle \ $denotes a state in which
only the $n$th atom is excited while all the other $N-1$ atoms are in their
ground states. From the equation of motion $i\partial \left| \Phi
_{N}(t)\right\rangle /\partial t=\mathcal{H}_{eff}\left| \Phi
_{N}(t)\right\rangle ,$ the time-dependent coefficients $C_{n}^{(N)}(t)$
must satisfy the differential equations
\begin{equation}
i\frac{\partial C_{n}^{(N)}(t)}{\partial t}=\xi _{N}C_{n}^{(N)}(t)-\eta
_{N}\sum_{l=1;l\neq n}^{N}C_{l}^{(N)}(t))  \label{17}
\end{equation}
for $n=1,2,...,N.$ Without loss of generality, we assume that at $t=0 $ the
atoms are in the state $|\phi _{1}\rangle =|e_{1}g_{2}...g_{N}\rangle \ $%
[i.e., under the initial conditions $C_{1}^{(N)}(0)=1,$ $%
C_{2}^{(N)}(0)=C_{3}^{(N)}(0)=...=C_{N}^{(N)}(0)=0].$ Then, the solution of
Eqs. (\ref{17}) can be found in the form
\begin{equation}
C_{1}^{(N)}(t)=\frac{1}{N}e^{-i(\xi _{N}+\eta _{N})t}\left( e^{iN\eta
_{N}t}+N-1\right) ,  \label{18}
\end{equation}
\begin{equation}
C_{2}^{(N)}(t)=C_{3}^{(N)}(t)=...=C_{N}^{(N)}(t)=\frac{1}{N}e^{-i(\xi
_{N}+\eta _{N})t}\left( e^{iN\eta _{N}t}-1\right) .  \label{19}
\end{equation}
As is evident from Eqs. (\ref{18}) and (\ref{19}), at $t\neq \frac{2k\pi }{%
N\eta _{N}}$\textbf{\ }$(k=0,1,2,...)$ all the coefficients $%
C_{n}^{(N)}(t)\neq 0$ $(n=1,2,...,N)$ and thus the state $\left| \Phi
_{N}(t)\right\rangle $ of Eq. (\ref{16}) is an $N$-atom $W$-class state. In
particular, omitting an unimportant common phase factor, states of the form
\begin{equation}
\left| \Psi _{N}\right\rangle =\frac{1}{N}[(N-2)|\phi _{1}\rangle
-2\sum_{n=2}^{N}|\phi _{n}\rangle ]  \label{20}
\end{equation}
are generated at
\begin{equation}
t=\frac{(2k+1)\pi }{N\eta _{N}}.  \label{21}
\end{equation}

\section{NUMERICAL ANALYSIS}

In order to verify the validity of the above theoretical result, we analyze
the system's dynamics by numerically solving the Schr\"{o}dinger equation
with the full Hamiltonian $H$ in Eq. (\ref{1}). Suppose that we aim at
generating the $N$-atom $W$-class state $\left| \Psi _{N}\right\rangle $
given in Eq. (\ref{20}). In Fig. 2 we plot the fidelity $F_{N}=\left\langle
\Psi _{N}\right| \rho _{N}(t)\left| \Psi _{N}\right\rangle $ of the atoms'
state $\rho _{N}(t)$ obtained from the numerical calculation with respect to
the state $\left| \Psi _{N}\right\rangle $ for various values of $N$ and the
parameters chosen as $\Delta /f=\nu /f=10.$ Figure 2(a) shows $F_{N}$ versus
dimensionless time $\tau =N\eta _{N}t,$ showing that $F_{N}\simeq 1$ at $%
\tau =(2k+1)\pi $ for any $N$ which is in agreement with the theoretical
result of Eq. (\ref{21}). Alternatively, we also plot $F_{N}$ versus another
dimensionless time $ft$ in Fig. 2(b), from which it follows that the time
for $F_{N}$ to reach $1$ is longer if $N$ is larger. This also agrees with
the theoretical result of Eq. (\ref{21}), because $N\eta _{N},$ with $\eta
_{N}$ defined by Eq. (\ref{15}), decreases with increasing $N. $ The slow
oscillation of $F_{N}$ in Fig. 2 is due to the ``hopping'' of atomic
excitation among the atoms due to the virtual excitation of the field modes,
as theoretically predicted by Eqs. (18) and (19) when the effective
Hamiltonian (\ref{13}) is used. As for the fast oscillation in the fidelity $%
F_{N}$, it results from the atom-cavity energy exchange based on the use of
the full Hamiltonian (\ref{1}).

To be more concrete, let us deal with a specific situation for $N=4$ and $%
t=\pi /4\eta _{4},$ i.e., the target $W$-class state is
\begin{equation}
\left| \Psi _{4}\right\rangle =\frac{1}{2}(|e_{1}g_{2}g_{3}g_{4}\rangle
-|g_{1}e_{2}g_{3}g_{4}\rangle -|g_{1}g_{2}e_{3}g_{4}\rangle
-|g_{1}g_{2}g_{3}e_{4}\rangle ).  \label{22}
\end{equation}
It is necessary to consider the influence of different detunings on the
fidelity $F_{4}=\left\langle \Psi _{4}\right| \rho _{4}(t)\left| \Psi
_{4}\right\rangle .$\textbf{\ }For $N=4$, the condition $\Delta /\nu <2$
should be satisfied according to Eq. (\ref{21}). Thus, we assume $\nu $ and $%
\Delta $ such that $\nu =10f$ and $2f<\Delta <12f$ for plotting the fidelity
$F_{4}$ against different $\Delta $ in Fig. 3, where the oscillatory
behavior is mainly caused by the energy exchange between the atoms and the
fields. The numerical result shows that the average fidelity becomes closer
to 1 for a larger detuning $\Delta .$ That is, a larger detuning suits the
scheme better under an ideal environment. However, a quantum system
interacts with the noisy environment inevitably, which induces unwanted
disturbance to the target entangled state. The decoherence originates from
physical factors such as the atomic spontaneous emission, the cavity decay
and the fiber decay. To account for these decoherence factors, we employ the
master equation for the density matrix $\rho $ of the whole system, which is
of the well-known form
\begin{eqnarray}
\dot{\rho} &=&-i[H,\rho ]-\sum_{l=1}^{N}\frac{\Gamma _{l}}{2}%
(S_{l}^{+}S_{l}^{-}\rho -2S_{l}^{-}\rho S_{l}^{+}+\rho S_{l}^{+}S_{l}^{-})
\nonumber \\
&&-\sum_{l+1}^{N}\frac{\gamma _{l}}{2}(a_{l}^{\dag }a_{l}\rho -2a_{l}\rho
a_{l}^{\dag }+\rho a_{l}^{\dag }a_{l})  \nonumber \\
&&-\frac{\kappa }{2}(b^{\dag }b\rho -2b\rho b^{\dag }+\rho b^{\dag }b),
\label{23}
\end{eqnarray}
where $\Gamma _{l}$ is the spontaneous emission rate from the excited state $%
|e\rangle $ to the ground state $|g\rangle $ of the $l$th atom, $\gamma _{l}$
is the decay rate of the $l$th cavity and $\kappa $ is the decay rate of the
optical star coupler. Assuming $\Gamma _{l}=\Gamma $ and $\gamma _{l}=\gamma
$ for simplicity, the dependence of the fidelity $F_{4}$ on $\Delta /f$ and $%
\Delta /\nu $ in Fig. 4 can be obtained by numerically solving the master
equation (\ref{23}). For $0.8<\Delta /\nu <2$ and $\Delta /f$ being a
constant, we find that the fidelity decreases quickly when $\Delta /\nu
\rightarrow 2,$ in which case the large detuning condition $|\Delta -\sqrt{4}%
\nu |\gg f$ is not satisfied. This implies that use of the effective
Hamiltonian in Eq. (\ref{13}) is not valid in this case. With the decreasing
of $\Delta /\nu ,$ the large detuning condition is getting satisfied and the
dynamics of the whole system will evolve in accordance with that governed by
the effective Hamiltonian. As seen from Fig. 4(a), the fidelity drops slowly
when $\Delta /\nu \leq 1$ and $\Gamma /f=0.01.$ This is because the
interaction time needed to achieve the target entangled state prolongs
according to Eq. (\ref{21}), causing more decoherence from atomic
spontaneous emission since the probability of population in the atomic
excited state is larger. The fidelity in Fig. 4(b) ((c)) keeps very high
even when $\Delta /\nu \leq 1$ and $\gamma /f=0.3\ (\kappa /f=0.3),$
revealing robustness of the fidelity against the cavity decay and fiber
decay due to the fact that the probabilities of real population in the
cavity and the fiber are negligible in the effective Hamiltonian (\ref{13})
that contains no interaction terms among those field modes. Keeping the
ratio $\Delta /\nu $ fixed in Fig. 4, we can consider the effect of
different $\Delta $ on the fidelity. From Fig. 4, we can see that the
increase of $\Delta $ will decrease the fidelity with an oscillatory
behavior. This is because although $\Delta $ is large the interaction time
prolongs with the increasing of $\Delta $ in accordance with Eq. (\ref{21}),
where the decoherence dominates the system dynamics. Therefore, the range $%
0.8<\Delta /\nu <1.2$ is the appropriate choice in our scheme. It also shows
that the fidelity is robust against the possible imprecision of the
atom-cavity detuning and the coupling strength between the bosonic modes.

Next, by choosing an appropriate values $\Delta /f=\nu /f=10$ in Figs. 5(a)
((b)), we plot the fidelity versus the ratios $\Gamma /f$ and $\gamma /f$ $%
(\Gamma /f$ and $\kappa /f).$ These figures indicate that the atomic
spontaneous emission dominates the reduction of fidelity, while the decay
rates of the photon leaking out off each cavity and the optical fiber
channels just slightly influence $F_{4},$ which is $0.93$ $(0.97)$ even when
$\gamma /f=0.3$ $(\kappa /f=0.3).$ Hence, the scheme is remarkably robust
against cavity and fiber decays, which can be understood by the virtual
excitation of all the bosonic field modes.

Finally, we briefly discuss the basic elements that may be a candidate for
further experiments. The requirements of our scheme are the two-level atoms
and the cavities resonantly connected by an optical fiber star coupler. The
single-mode integrated optical fiber $1\times N$ star coupler used as a
distributed strain sensor in a white-light interferometer has been reported
\cite{ao-lby68} and realized by using a 2D arrangement, by using the two
confocal arrays of the radial waveguides which performs with an efficiency
100\% under ideal conditions when the waveguides' mutual coupling strange is
strong \cite{ieee-cd41}. A near-perfect fiber-cavity coupling with an
efficiency larger than 99.9\% can be realized using fiber-taper coupling to
high-Q silica microspheres \cite{ieee-kjg00}. The atomic configuration can
be achieved with cesium: state $|g\rangle $ corresponds to $\{F=4,m=3\}$
hyperfine state of $6^{2}S_{1/2}$ electronic ground state and state $%
|e\rangle $ corresponds to $\{F=4,m=3\}$ hyperfine state of $6^{2}P_{1/2}$
electronic state. Each single atom can be made localized at a fixed position
in each cavity with high Q for a long time \cite
{PRL-89-103001-2002,PRA-71-13817-2005,PRL-87-037902-2001}. In recent
experiments \cite{sm-pra817,jr-pra806}, the parameters $f=2\pi \times 750$
MHz$,\ \Gamma =2\pi \times 2.62$ MHz and $\gamma =2\pi \times 3.5$ MHz with
the wavelength in the region $630\sim 850$ nm is predicted achievable. The
optical fiber decay at a $852$ nm wavelength is about $2.2$ dB/km \cite
{pra-fd04}, which corresponds to fiber decay rate $0.152$ MHz. By
substituting these experimental parameters into Eq. (\ref{23}), we obtain a
fidelity higher than $0.9,$ making our scheme possible to be realized in
experiment. The generation of $W$-class states involving more atoms is also
efficient by changing the corresponding experimental parameters.

\section{CONCLUSION}

We have considered a model consisting of any $N\geq 3$ identical two-level
atoms trapped in $N$ spatially separated cavities. Each cavity has one
active mode which is off-resonant with the atomic transition but resonant
with a single mode of an integrated optical $1\times N$ star coupler (see
Fig. 1), so all the atoms are indirectly coupled to each other even though
they are far apart. We deduce an effective Hamiltonian in the large
atom-cavity mode detuning regime and use it to theoretically study the
dynamics of the atoms' system under the initial condition that only one atom
is excited while all the cavities and the fiber are empty. The theoretical
result shows that as the system evolves the atoms generally appear in an $N$%
-atom $W$-class state. Of interest are the multiatom entangled states of the
form in Eq. (\ref{20}) which are generated periodically at time moments
determined by Eq. (\ref{21}). The proposed scheme for generating $N$-atom $W$%
-class states does not require any external classical laser pulses and is
insensitive to the cavity decay rate and the rate of photon leakage from the
fiber because during the whole evolution no real bosons are to be created
due to the large detuning between the atomic transition and the cavity mode.
We have also carried out numerical calculations taking into account the
effects of decoherence caused by various dissipation mechanisms. The
numerical result agrees well with the theoretical one if the atom-cavity
detuning is large enough, confirming the validity of the effective
Hamiltonian with respect to the full one. Therefore, the present
entanglement generation scheme proves to be perspective for wide
applications in the scalable distributed quantum networks.

\section{Acknowledgments}

M. L and Y. X were supported by the National Natural Science Foundation of
China under grant no. 11047122 and no. 11105030, and China Postdoctoral
Science Foundation under grant no. 20100471450. N. B. A. was funded by
Vietnam National Foundation for Science and Technology Development
(NAFOSTED) under grant no. 103.99-2011.26.

\newpage
\begin{figure}[tbp]
\centering\subfigure[]{\label{Fig.sub.a}\includegraphics[width=5cm]{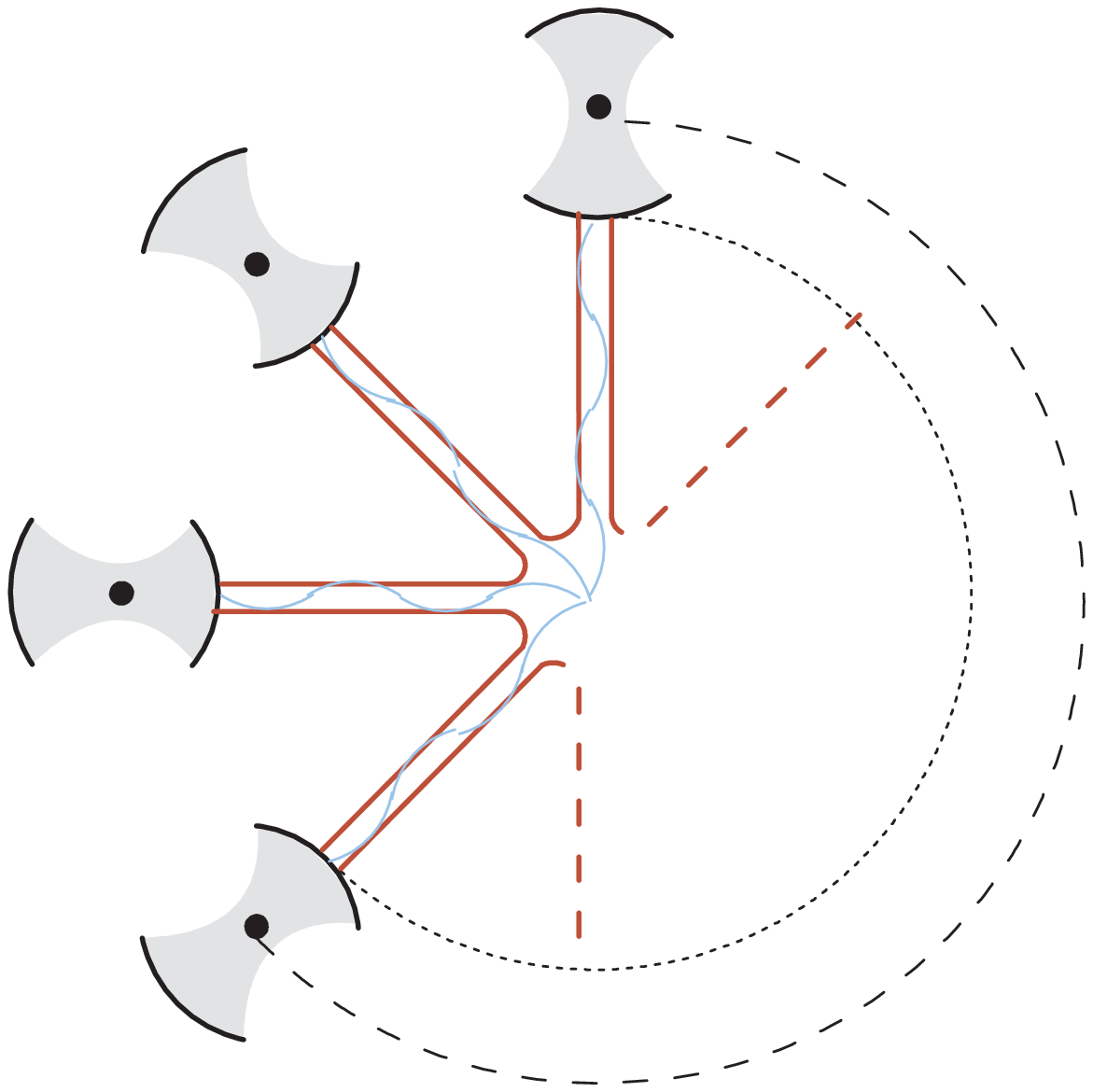}%
}
\subfigure[]{\label{Fig.sub.b}\includegraphics[width=5cm]{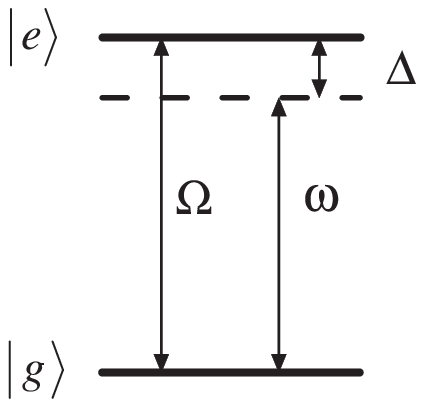}}
\caption{(color online) (a) Experimental setup. The block dots
denote the atoms, which are trapped in $N$ distant cavities, and
these cavities are connected by a $1\times N$ single-mode integrated
optical star coupler. (b) Level configuration for each atom.}
\label{Fig.1}
\end{figure}

\begin{figure}[tbp]
\centering\subfigure[]{\label{Fig.sub.a}%
\includegraphics[width=12cm]{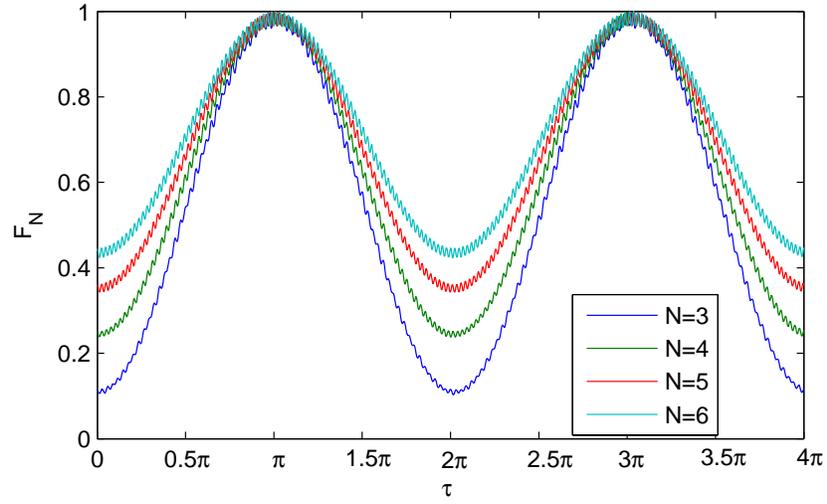}} \subfigure[]{\label{Fig.sub.b}%
\includegraphics[width=12cm]{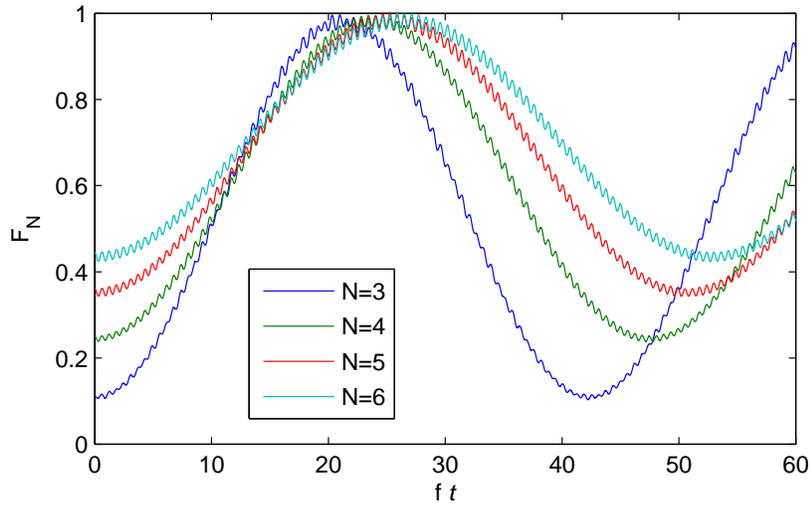}}
\caption{(color online) The fidelity $F_N$ versus dimensionless time (a) $%
\tau=N\eta_Nt$ and (b) $ft$, with $\Delta/f=\Delta/\nu=10$ for $N=3,\ 4,\ 5$
and $6$. }
\label{Fig.2}
\end{figure}

\begin{figure}[tbp]
\scalebox{0.8}{\includegraphics {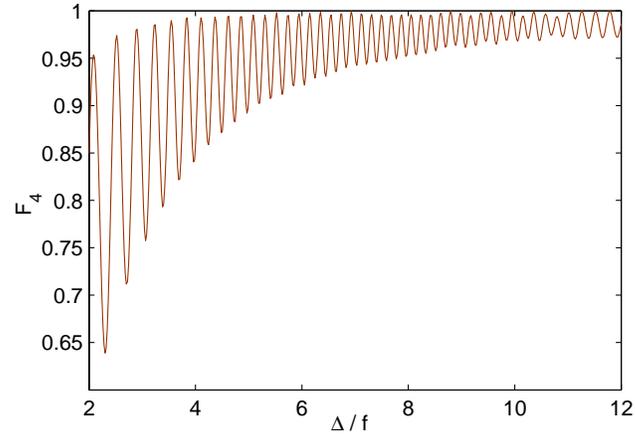}}
\caption{(color online) The fidelity $F_4$ at $t=\pi /4\eta _{4}$ versus $%
\Delta/f$ when $\nu/f=10$.}
\label{Fig.3}
\end{figure}

\begin{figure}[tbp]
\centering\subfigure[]{\label{Fig.sub.a}%
\includegraphics[width=10cm]{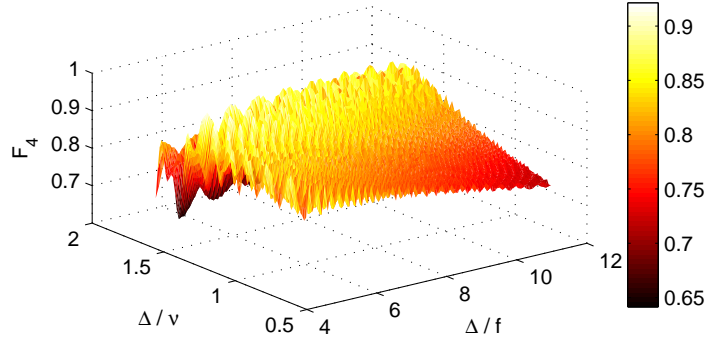}} \subfigure[]{\label{Fig.sub.b}%
\includegraphics[width=10cm]{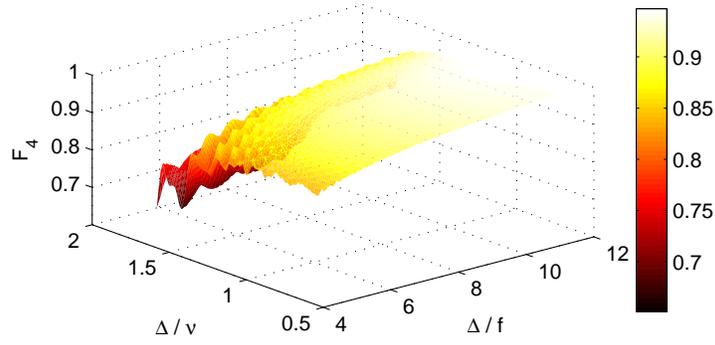}} \subfigure[]{\label{Fig.sub.b}%
\includegraphics[width=10cm]{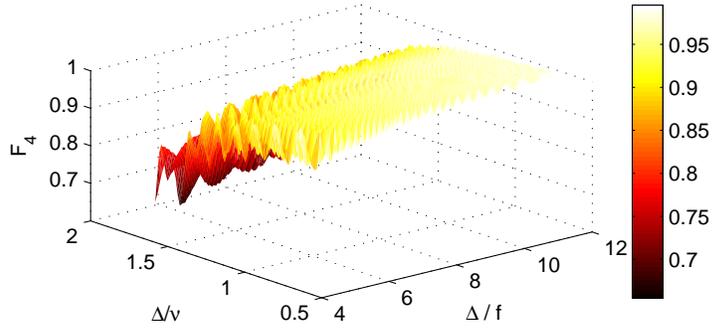}}
\caption{(color online) The fidelity $F_4$ at $t=\pi /4\eta _{4}$ versus $%
\Delta/f$ and $\Delta/\nu$ when (a) $\Gamma/f=0.01$ and $\gamma=\kappa=0$;
(b) $\Gamma=\kappa=0$ and $\gamma/f=0.3$ and, (c) $\Gamma=\gamma=0$ and $%
\kappa/f=0.3$.}
\label{Fig.4}
\end{figure}

\begin{figure}[tbp]
\centering\subfigure[]{\label{Fig.sub.a}\includegraphics[width=8cm]{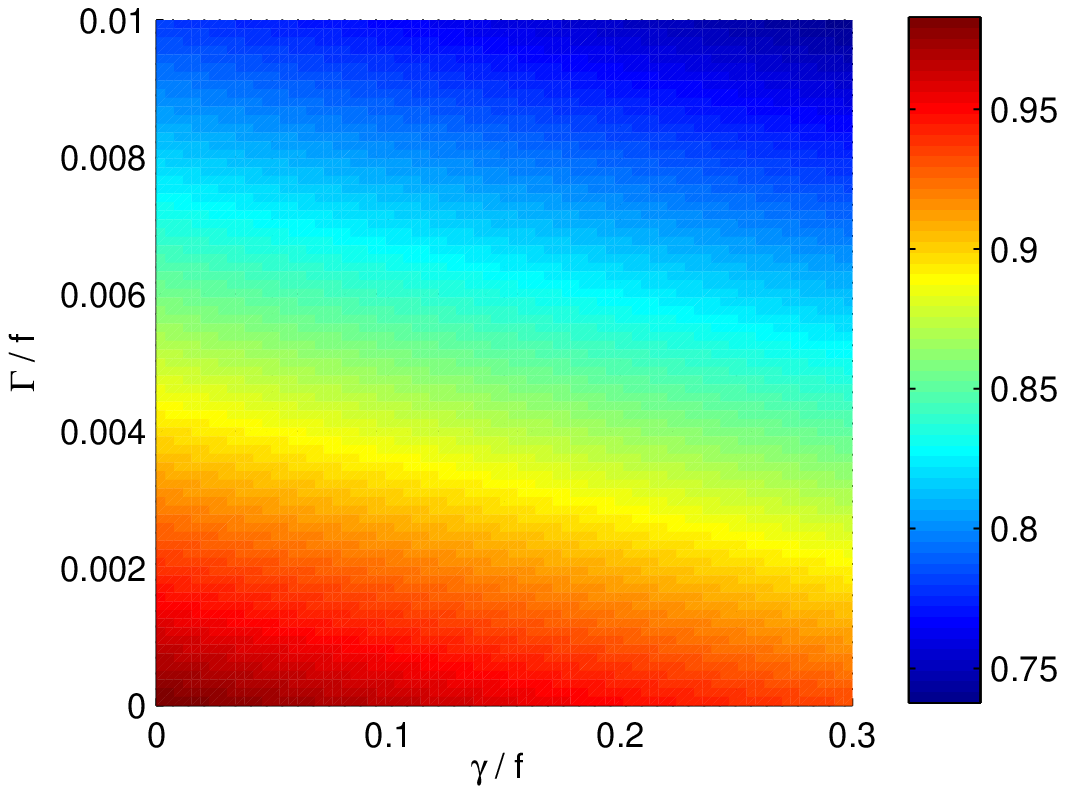}%
}
\subfigure[]{\label{Fig.sub.b}\includegraphics[width=8cm]{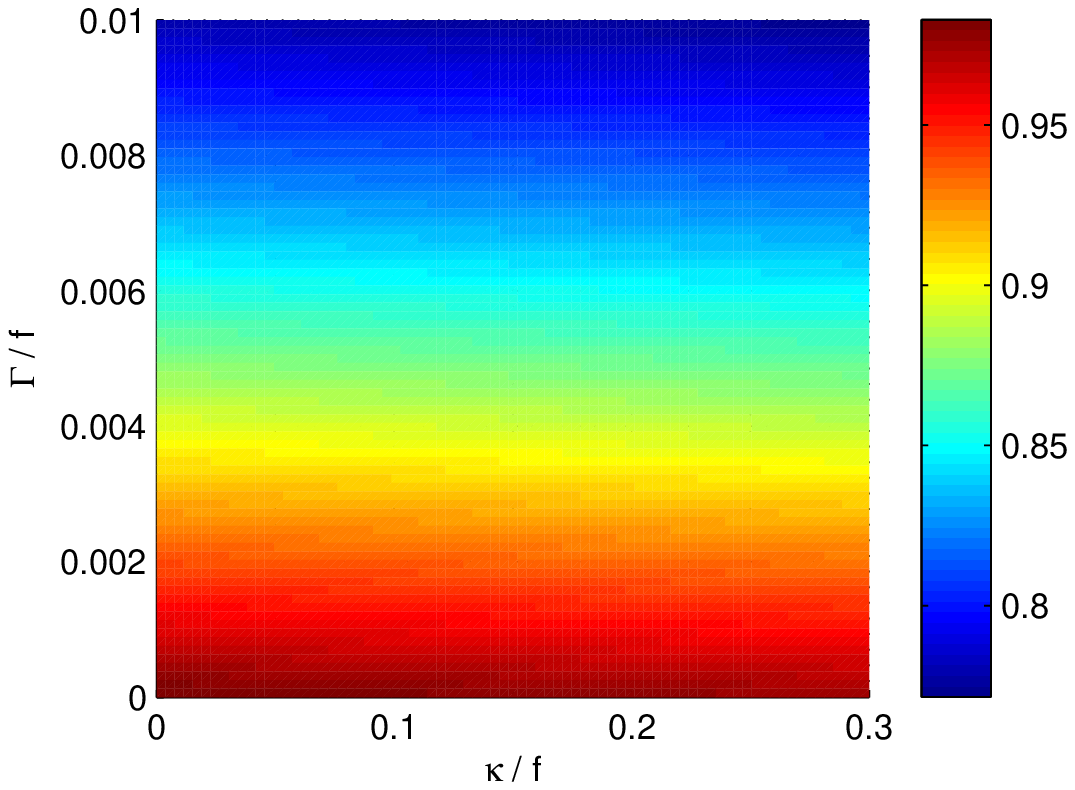}}
\caption{(color online) Density plots of the fidelity $F_4$ at
$t=\pi /4\eta _{4}$ versus (a) $\Gamma/f$ and $\gamma/f$ and (b)
$\Gamma/f$ and $\kappa/f$. } \label{Fig.5}
\end{figure}

\end{document}